\documentclass[onecolumn,aps,floatfix]{revtex4-2}
\usepackage[centertags]{amsmath}
\usepackage{amsfonts}
\usepackage{amssymb}
\usepackage{amsthm}
\usepackage{mathptmx}
\usepackage{changes}
\usepackage{newlfont}
\usepackage{stmaryrd}
\usepackage{mathrsfs}
\usepackage{euscript}
\usepackage{siunitx}
\usepackage{comment}
\usepackage{algorithm2e}
\usepackage{graphicx,subcaption}
\usepackage{enumitem}
\usepackage{natbib} 
\usepackage{braket}
\usepackage{bbm}
\usepackage{graphicx}
\usepackage{xcolor}
\usepackage{orcidlink}
\usepackage{floatrow}
\usepackage{caption}
\usepackage{hyperref}
\usepackage{subcaption}
\usepackage{hhline}
\usepackage{hyperref}
\usepackage{cleveref}
\usepackage{amsmath}
\usepackage{bm}
\usepackage{tikz}
\usetikzlibrary{arrows.meta, positioning, decorations.pathmorphing}
\usepackage{pgfplots}
\usepgfplotslibrary{fillbetween}
\usepackage{xcolor}
\pgfplotsset{compat=1.18}
\makeatletter \def\switch@array{}\makeatother

\def\bea{\begin{eqnarray}}
\def\eea{\end{eqnarray}}
\def\ba{\begin{array}}
\def\ea{\end{array}}

\theoremstyle{plain}

\theoremstyle{definition}

\theoremstyle{remark}

\numberwithin{equation}{section}

 \let\la=\lambda

\newcommand{\opunit}{\text{1}\kern-0.22em\text{l}}

\DeclareMathAlphabet{\mathpzc}{OT1}{pzc}{m}{it}

\newcommand{\fig}{Fig.}

\newcommand{\id}{\textrm{d}}

\usepackage{floatrow}
\usepackage{caption}
\DeclareCaptionJustification{justified}{\leftskip=0pt \rightskip=0pt \parfillskip=0pt plus 1fil}

\numberwithin{equation}{section}

\usepackage{xcolor,soul}

\usepackage{makeidx}
\usepackage{comment}
\usetikzlibrary{decorations.pathmorphing}
\def\bea{\begin{eqnarray}}
\def\eea{\end{eqnarray}}
\def\ba{\begin{array}}
\def\ea{\end{array}}

\def\bea{\begin{eqnarray}}
\def\eea{\end{eqnarray}}
\def\ba{\begin{array}}
\def\ea{\end{array}}

\def\la{\langle}
\def\ra{\rangle}

\definecolor{dgreen}{rgb}{0,0.7,0}

\DeclareUnicodeCharacter{202F}{\,}
\begin{document}
\title{Specific heat of thermally driven chains}

\author{Michiel Gautama  \orcidlink{0009-0009-6154-7468}} \author{Faezeh Khodabandehlou \orcidlink{0000-0001-8114-6105}} \author{Christian Maes \orcidlink{0000-0002-0188-697X}} \author{Ion Santra \orcidlink{0000-0002-9772-2880}} \affiliation{Department of Physics and Astronomy, KU Leuven}

\begin{abstract}
We investigate the thermal responses of a harmonic oscillator chain coupled at its boundaries to heat baths held at different temperatures. This setup sustains a steady energy flux, continuously dissipating heat into both reservoirs. By introducing slow variations in the bath temperatures, we quantify the resulting excess heat currents and thereby obtain the nonequilibrium heat capacity matrix at fixed but arbitrary temperature differences.  We demonstrate the existence of a well-defined thermodynamic limit for long chains. The specific heat associated with energy exchanges with a single bath depends on the (difference in) friction coefficients governing the system–bath couplings.  That thermokinetic effect is typical for nonequilibrium response.  When the couplings with the thermal baths acquire temperature dependence, the specific heat correspondingly inherits a nontrivial temperature dependence, in sharp contrast with equilibrium.
Our results provide the first explicit determination of specific heat(s) in a locally interacting, spatially extended driven system. Beyond its exact solvability, the model may offer a natural nonequilibrium extension of the Dulong–Petit law, capturing the high-temperature behavior of driven molecules.
\end{abstract}

\maketitle

\section{Introduction}

Specific heats constitute important markers of the number and nature of degrees of freedom in the system. In that sense, calorimetry has been crucial in the statistical mechanical understanding of thermodynamics.  
As a historical example, the transition from classical to quantum mechanics has been accompanied by detailed studies of low-temperature heat capacities \cite{DulongPetit1819,Einstein1907, Debye1912}. For soft matter systems as well, calorimetry has been an essential tool, for instance in characterizing phase transitions~\cite{yeomans1992statistical,iannacchione1992calorimetric,wunderlich1997heat}.\\ In the present paper, we turn to an exactly solvable yet paradigmatic nonequilibrium system, investigating the quasistatic thermal response of a boundary driven spatially-extended system of coupled harmonic oscillators.  That heat-conducting harmonic chain is Hamiltonian in the bulk, while the boundaries are subject to Langevin forces that represent contact with different heat baths \cite{rieder1967properties,nakazawa1970lattice}. Nonequilibrium driving is introduced through a temperature difference between the reservoirs, leading to a steady heat current across the chain. While this setup has been extensively investigated as a prototype of heat conduction, both for harmonic and anharmonic chains, providing key insights into anomalous transport in low-dimensional crystals \cite{dhar2001heat,lepri2003thermal,lepri2005studies,aoki2006energy,dhar2008heat,lukkarinen2008anomalous,Monasterio2019}, we shift the focus from transport to calorimetry, computing the specific heat matrix. We use the method first explained in \cite{epl,cejp,jir,calo} and applied to small theoretical model systems; see e.g. \cite{jchemphys} and references therein. \\

Our central result is obtaining the heat capacity matrix, a $2\times 2$ response matrix that quantifies how the heat currents into each reservoir change under slow variations of the bath temperatures. Owing to the linearity of the dynamics, this matrix can be computed, retaining its full dependence on system size, temperatures, and coupling strengths. We establish the existence of a well-defined thermodynamic limit and analyze its structure in detail as the chain length goes to infinity (and the harmonic chain remains conducting). To compare with equilibrium, we isolate the response in heat flux to one thermal bath and we find that it is temperature-independent (as in equilibrium); yet it depends on the friction coefficients (in contrast with equilibrium). 
As a further consequence, when those friction coefficients (effectively) depend on temperature(s) themselves, as easily happens when far from equilibrium and for certain bath-couplings, the specific (heats) also start to depend on temperature (unseen in thermodynamic equilibrium). The lesson is that specific heat (outside equilibrium) is not just a material constant (where temperature excites the for that scale relevant degrees of freedom); it also depends on the functioning of the material, here in terms of its coupling with and the heat conduction between two thermal baths.\\

Beyond providing, to our knowledge, the first exact computation of specific heats in a locally interacting driven macroscopic system, our results suggest a broader perspective. In the spirit of the classical Dulong–Petit law~\cite{DulongPetit1819}, which captures universal high-temperature behavior via harmonic approximations~\cite{kittel2018introduction}, our findings point toward a nonequilibrium extension: whenever interactions can be effectively linearized, boundary-driven systems should exhibit analogous universal features. In particular, we predict that the molar heat capacity of a heat-conducting gas becomes constant when the excess heat flux is defined by variations in a single reservoir temperature.\\

The rest of the paper is organized as follows: we introduce the general setup for computing the nonequilibrium specific heat of a driven extended system in Sec.~\ref{sec:setup}. We use that in Sec.~\ref{sec:harmonic1} to compute the specific heat of a one-dimensional harmonic chain driven by thermal baths at the two ends. We discuss the effect of driving by a particle/soft-matter bath at one end in Sec.~\ref{sec:sm}. In Sec.~\ref{sec:mech} we discuss the quasistatic response to a mechanical perturbation. We conclude with Sec.~\ref{sec:conc}. 

\section{Setup}\label{sec:setup}

\begin{figure}[H]
    \centering
   \begin{tikzpicture}[scale=0.7, transform shape, font=\small]
            \def\N{5}        
            \def\dx{1.5cm}   
            
            \draw[fill=cyan!20, draw=blue] (-1.5cm,-1cm) rectangle (0.5cm,1cm);
            \node at (-0.5cm,0.5cm) {\scalebox{1.5}{$T_\ell$}};
            
            \draw[fill=purple!20, draw=red] (6.5cm,-1cm) rectangle (8.5cm,1cm);
            \node at (7.5cm,0.5cm) {\scalebox{1.5}{$T_r$}};
            
            \draw[fill=black] (-0.5cm,0) circle (0.08cm);
            \draw[fill=black] (7.5cm,0) circle (0.08cm);

            \draw[fill=black] (1.5,0) circle (0.08cm);
            \draw[decorate, decoration={zigzag, segment length=6pt, amplitude=3pt}](-0.5,0) -- (1.5,0);
            \draw[decorate, decoration={zigzag, segment length=6pt, amplitude=3pt}](-0.5,0) -- (-0.5,-0.75);
            \draw[ultra thick](-1,-0.75) -- (0,-0.75);
            \draw[fill=black] (5.5,0) circle (0.08cm);
            \draw[decorate, decoration={zigzag, segment length=6pt, amplitude=3pt}] 
                (5.5,0) -- (7.5,0);
            \draw[decorate, decoration={zigzag, segment length=6pt, amplitude=3pt}](7.5,0) -- (7.5,-0.75);
            \draw[ultra thick](7,-0.75) -- (8,-0.75);
            
            \draw[decorate, decoration={zigzag, segment length=6pt, amplitude=3pt}] 
                (1.5,0) -- (3,0);
            \draw[decorate, decoration={zigzag, segment length=6pt, amplitude=3pt}] 
                (4,0) -- (5.5,0);
            
            \node[below] at (3.5,0) {$\dots$};
   \end{tikzpicture}
    \caption{\small{Chain of $N$ oscillators coupled to two heat baths.}}
    \label{chain}
\end{figure}
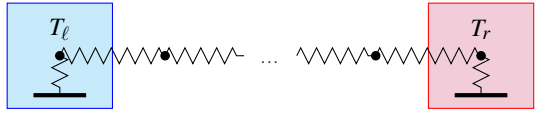

We start from a microscopic model of a spatially--extended system that is coupled to two heat baths. For that purpose, we consider a chain of $N$ oscillators $(q_j,p_j)$, each  of mass $m=1$, with displacements $q_j$ and momenta $p_j$, $j=1,\dots,N$.  We are mostly interested in large $N$.  The bulk dynamics is Hamiltonian where $H(q,p)=\sum_{i=1}^N \frac{p_i^2}{2} + U(q_1,\dots,q_N)$ contains a self-energy and an arbitrary local interaction potential $U(q)
=
\sum_{i=1}^N U_i(q_i)
+
\sum_{i<j} U_{\text{int}}(q_i-q_j)$. The chain is coupled at its boundaries to Langevin heat baths at temperatures $T_1 = T_\ell$ and $T_N=T_r$; see \fig \ref{chain}.
The equations of motion (with $k_B=1$) are
\begin{align}
\dot q_i &= p_i, \quad i=1,\dots,N,
\qquad\dot p_i = -\partial_{q_i} U(q), \qquad i=2,\dots,N-1,
\label{eq:pdot_bulk}
\\
\dot p_{1} &= -\partial_{q_1} U(q)- \gamma_\ell p_1
+\sqrt{2\gamma_\ell T_\ell}\,\eta_\ell(t),\\
\dot p_{N} &= -\partial_{q_N} U(q)- \gamma_r p_N
+\sqrt{2\gamma_r T_r}\,\eta_r(t)
\label{eq:pdot}
\end{align}
where $\eta_\ell,\eta_r$ are independent Gaussian white noises with zero mean, and $\gamma_\ell,\gamma_r$ are the associated friction coefficients. The backward generator acting on functions $f(q,p)$ is 
\begin{align}
L f
&=
\sum_{i=1}^N p_i\,\partial_{q_i} f
-
\sum_{i=1}^N \partial_{q_i}U(q)\,\partial_{p_i} f
-\gamma_\ell p_1\,\partial_{p_1} f
+\gamma_\ell T_\ell\, \partial_{p_1}^2 f
-\gamma_r p_N\,\partial_{p_N} f
+\gamma_r T_r\, \partial_{p_N}^2 f 
\label{eq:backward_generator_general}
\end{align}
For fixed bath temperatures $(T_\ell\neq T_r)$, the system reaches a nonequilibrium stationary condition described by density $\rho^s$ (with respect to $\id X =\id p\,\id q$), and averages are denoted with $\langle\cdot\rangle^s$.  There is a steady current flowing from the hotter bath to the colder bath, \cite{rieder1967properties,dhar2001heat,maes2003heat}.  In that way, the stationary heat-conducting chain is an open thermally driven, hence steady nonequilibrium, macroscopic system.  We are interested in the extra heat that flows from the baths to the system when a parameter $\lambda$ (such as $T_r$ or $T_\ell$) (slowly) changes over a (long) time-interval; that is called quasistatic thermal response.\\

\begin{figure}[H]
    \centering
    \scalebox{0.60}{
    \begin{tikzpicture}
    \begin{axis}[
            axis lines = middle, 
            xlabel = {Time $t$}, 
            ylabel = {Power $P$}, 
            xmin = -50, xmax = 300, 
            ymin = -10, ymax = 60, 
            width = 16cm,
            height = 8cm,
            x axis line style={ultra thick}, 
            y axis line style={ultra thick},
            xtick ={\empty},
            ytick = {\empty},
            samples=200,
            x label style={anchor=west, font = \fontsize{13pt}{13pt}\selectfont}, 
            y label style={anchor=south, font = \fontsize{13pt}{13pt}\selectfont},
            legend style={at={(0.75,0.7)},anchor=south}
        ]
    
            \addplot[color = blue, domain = -50:0, ultra thick]{40};
            \addlegendentry{\scalebox{1.5}{$P_\ell$}}
            \addplot[color = purple, domain = -30:0, ultra thick]{40};
            \addlegendentry{\scalebox{1.5}{$P_r$}}

            \addplot[name path = A1, color = purple, ultra thick, domain = 0:150, samples=200]{
                (sqrt((40-10)^2 - x^2/25) + 2.5*sin(deg(x/5))) * exp(-0.03*x/5)
                + 5 + 5*exp(-0.08*x/5)
                + 23*exp(-((x - 50)^2)/100)
                - 23*exp(-((x - 130)^2)/80)
                + 7.5*exp(-((x - 150)^2)/50)
            };
    
            \addplot[name path = A2, color = purple, ultra thick, domain = 150:300, samples=200]{
                12.06 + 10*exp(-((x - 170)^2)/50)
            };
    
            \addplot[name path=E1, color=blue, ultra thick, domain=0:170, samples=200]{
              40
            + 12*exp(-((x -  30)^2)/180)      
            + 24*exp(-((x -  80)^2)/200)      
            - 16*exp(-((x - 100)^2)/60)       
            + 17*exp(-((x - 135)^2)/80)       
            +  5*sin(deg(x/11))*exp(-x/120)   
            - 27.94*(1 - exp(-x/18))          
            };
    
            \addplot[name path=E2, color=blue, ultra thick, domain=170:300, samples=200]{
            12.06
            };

            \addplot[name path = B, color = gray, dashed, ultra thick, domain = -50:200]{12.06};
    
            \addplot[name path = C, color = black, ultra thin, domain = 0:300]{0};
    
            \addplot[name path = D, color = purple, ultra thick, domain = 199:300]{12.06};
    
            \addplot[gray!20] fill between[of= B and C, soft clip={domain=0:290}];
            \addplot[gray!20] fill between[of= D and C, soft clip={domain=0:290}];
            \addplot[purple!20] fill between[of= A1 and B, soft clip={domain=0:150}];
            \addplot[purple!20] fill between[of= A2 and B, soft clip={domain=150:200}];
            \addplot[cyan!20] fill between[of= E1 and B, soft clip={domain=0:170}];
            \addplot[cyan!20] fill between[of= E2 and B, soft clip={domain=170:200}]; 
    
            \node [above] (2) at (80,50) {\scalebox{1.5}{$ \delta Q_r^{\text{exc}}$}};
            \node [above] (3) at (135,35) {\scalebox{1.5}{$ \delta Q_\ell^{\text{exc}}$}};
            \node [above] at (-30,-10) {\scalebox{1.5}{$T_\ell$}};
            \node [above] at (30, -10) {\scalebox{1.5}{$T_\ell+dT_\ell$}};
            \node (1) [above] at (45,25.5) {};
            \node (4) [above] at (80,25) {};
            \draw[->, line width = 0.7 mm, color= purple]  (1) [out=70, in=-150] to  (2);
            \draw[->, line width = 0.7 mm, color= blue]  (4) [out=70, in=-150] to  (3);
    
            \draw (50,0) -- node[fill=white, ultra thick, rotate=0,inner xsep=-4 pt, inner ysep = -5]{\scalebox{1.3}{//}} (60,0);
            \node [above] at (-8, -7) {\scalebox{1.2}{0}};
        \end{axis}
        \end{tikzpicture}
    }
    
    \caption{A sketch of the nonequilibrium heat fluxes $P_\ell,P_r$ as a function of time for a process where the left bath temperature is changed $T_\ell\to T_\ell+dT_\ell$. The heat fluxes start from a steady state at temperature $T_\ell$, where $P_\ell+P_r=0$, to a new steady state. We calculate the excess heats as the area between the intermediate response and the eventual steady state.}
    \label{cartoon}
\end{figure}
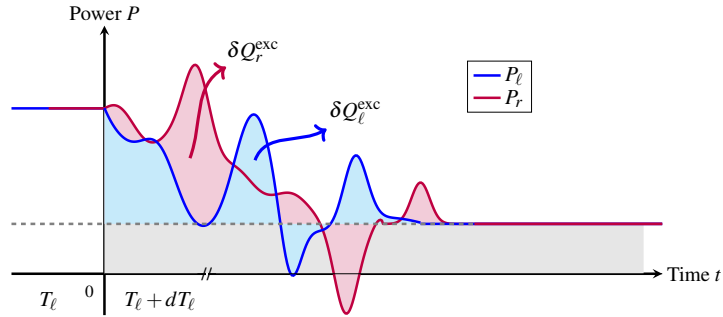

To understand the thermal response for that heat-conducting chain, we need the theory of quasistatic perturbations; see \cite{berryo}.  The observable (that is being measured and for which we want to know the response) is the excess heat $\delta Q_r^\text{exc}$ and $\delta Q_\ell^\text{exc}$, as represented in the cartoon Fig.~\ref{cartoon}.  Remark that even before perturbing, the steady system receives all the time heat (positive and negative) from both right and left reservoirs; by $\delta Q_r^\text{exc}$ and $\delta Q_\ell^\text{exc}$, we mean the extra heat to the system which is entirely due to the perturbation.   It allows to define the heat capacity matrix with elements
\begin{equation}\label{hcm}
C_{\alpha \alpha'} = \frac{\delta Q_{\alpha'}^\text{exc}}{\id T_\alpha},\qquad \alpha,\alpha' \in \{\ell, r\}.
\end{equation}
Here the first index $\alpha$ indicates the bath whose temperature is changed by $\id T_\alpha$, while the second index $\alpha'$ labels the bath whose excess heat $\delta Q_{\alpha'}^\text{exc}$ is measured.

We add some of the mathematical formalism; we refer to \cite{jir,cejp,epl,jchemphys} for more details.  It starts by measuring the heat currents $\dot Q_{\ell}$ and $\dot Q_{r}$ injected by the left and right reservoirs, respectively,
\begin{align}
\dot Q_{\ell,r}
&=
\Big[-\gamma_{\ell,r} \,p_{1,N} + \sqrt{2\gamma_{\ell,r} T_{\ell,r}}\,\eta_{\ell,r}(t)\Big] \circ p_{1,N}\end{align}
where $\circ$ denotes the Stratonovich product. By averaging over the noise and fixing the state $X=(q,p)=(q_1,\dots,q_N,p_1,\dots,p_N)^\dagger$, we get the expected heat fluxes from the left and right thermal baths,
\begin{align}
P_{\ell}(q,p)
=
\gamma_{\ell}\left(T_{\ell} - p_{1}^2\right),\qquad P_{r}(q,p)
&=
\gamma_{r}\left(T_{r} - p_{N}^2\right).
\label{eq:PLR}
\end{align}
The associated (instantaneous) excess heat flux is
\begin{align}
f_{\alpha'}(X)=P_{\alpha'}(X)-\langle P_{\alpha'}\rangle_\lambda^s,
\label{eq:f}
\end{align}
where we have subtracted the stationary average and indicated the dependence on the externally controlled parameter $\lambda$, such as a temperature or an interaction parameter. Its slow variation provides the quasistatic perturbation, and we are interested in the corresponding quasistatic response  $R_{\lambda\alpha'}$ of the excess heat flux \eqref{eq:f}.  The result, as explained in \cite{mathnernst,jchemphys,berryo}, is
\begin{equation}
    R_{\lambda\alpha'} 
    =\la\partial_\lambda V_{\alpha'}\ra^s_{\lambda}\label{eq:qsresp}
\end{equation}
where $V_{\alpha'}$ is the quasipotential, defined as the unique solution of the Poisson equation, \cite{pois},
\begin{align}
     L_{\lambda} V_{\alpha'} +f_{\alpha'}  = 0, \text{  with  } \la V_{\alpha'} \ra^s_\lambda=0 \label{eq:poisson}
\end{align}
for the generator \eqref{eq:backward_generator_general} (where the subscript indicates the stimulated parameter).  It is easy to check that in equilibrium, $T_\ell = T_r$, the quasipotential is essentially the energy, $ \langle H \rangle_\lambda^\text{eq}-H$.  It turns out that the heat capacity matrix \eqref{hcm} can be computed as 
\begin{equation}\label{hca}
C_{\alpha\alpha'} = \la\partial_{T_\alpha} V_{\alpha'}\ra^s_\lambda
\end{equation}

 In the following section, we apply that scheme to a harmonic chain pinned at the left and right boundaries, where the parameter $\lambda$ is the left or right temperature for obtaining the nonequilibrium heat capacity \eqref{hcm} via \eqref{hca}.
\section{Harmonic chain}
We take the case where $U_{\text{int}}(x)=kx^2/2$, and $U_i=0$ for $i=2,\cdots,N-1$ while $U_1(q_1)=\kappa q_1^2/2, U_N(q_N)=\kappa q_N^2/2 $.

 \subsection{The algebra}\label{sec:harmonic1}
 It is convenient to write the equations of motion in matrix notation,
\begin{equation}
\dot{X} = A X + \boldsymbol{\xi},
\end{equation}
where $A$ is the drift matrix and $\boldsymbol{\xi}$ is the noise vector, given by
\begin{equation}
A =
\begin{pmatrix}
0 & I \\[1mm]
-kK & -\Gamma
\end{pmatrix},
\qquad
\xi_i = \left(\sqrt{2\gamma_\ell T_\ell}\,\eta_\ell\, \delta_{i,N+1}
+ \sqrt{2\gamma_r  T_r}\,\eta_r\, \delta_{i,2N}\right).
\label{eq:A}
\end{equation}
The matrices $K$ and $\Gamma$ have elements
\begin{equation}
K_{ij} = 2\delta_{ij} - \delta_{i,j+1} - \delta_{i,j-1}
+ \left(\frac{\kappa}{k} - 1\right)(\delta_{i1}\delta_{j1} + \delta_{iN}\delta_{jN}),\quad \Gamma_{ij} = \gamma_\ell \delta_{i1}\delta_{j1} + \gamma_r \delta_{iN}\delta_{jN}.
\label{eq:K}
\end{equation}
The stochastic forcing acts only on the boundary momenta. Accordingly, the noise covariance matrix takes the form
\begin{equation}
D =
\begin{pmatrix}
0 & 0 \\
0 & \Theta
\end{pmatrix},
\qquad
\Theta_{ij} = 2\left(\gamma_\ell T_\ell\, \delta_{i1}\delta_{j1}
+ \gamma_r T_r\, \delta_{iN}\delta_{jN}\right).
\label{eq:D}
\end{equation}
The corresponding backward generator in this case becomes
\begin{equation}
L = (AX) \cdot \nabla + \frac{1}{2} \sum_{i,j} D_{ij}\,\partial_i \partial_j.
\label{eq:L}
\end{equation}
The stationary distribution $\rho^s(X)$, \cite{rieder1967properties}, is
\begin{equation}
\rho^s(X) \propto
\exp\!\left[-\frac{1}{2} X^{\dagger}\Sigma^{-1}X\right],\quad \text{where }A\Sigma + \Sigma A^{\dagger} = -D.
\label{eq:Sigma}
\end{equation} 
It is useful to note that, in the nonequilibrium stationary state, the bulk kinetic temperature profile is flat~\cite{lepri2003thermal}. The form of the bulk temperature for unequal dissipations, obtained via using the nonequilibrium Greens function methods~\cite{dhar2006heat,santra2022activity} are given by, 

\begin{align}\label{eq:bulkT}
T_{\text{bulk}}=\mathcal{A}_\ell T_\ell+\mathcal{A}_r T_r, \qquad \mathcal A_{\ell}=\frac{\gamma_r}{\gamma_r+\gamma_\ell}+\frac{(\gamma_\ell-\gamma_r)(k/\kappa)^2}{(\gamma_r+\gamma_\ell)(1+4\gamma_r\gamma_\ell/\kappa)^{1/2}}, \quad \mathcal A_r=\mathcal A_{\ell}\Big|_{\gamma_\ell\leftrightarrow\gamma_r}
\end{align}
 The centered source term defined in \eqref{eq:f} can be written as,
\begin{equation}
f_{\alpha'}(X) =X^\dagger F^{(\alpha')} X - \mathrm{Tr}(F^{(\alpha')}\Sigma),\quad\text{with  }F^{(\alpha')} =
\begin{pmatrix}
0 & 0 \\
0 & -H^{(\alpha')}
\end{pmatrix},
\label{eq:F}
\end{equation}
where $H^{(\ell)}_{ij} = \gamma_\ell \delta_{i1}\delta_{j1}$,  $H^{(r)}_{ij} =\gamma_r \delta_{iN}\delta_{jN},$  and $\Sigma$ is solved from \eqref{eq:Sigma}.
Clearly, since $f_{\alpha'}$ is quadratic and $L$ is linear, the quasipotential $V_{\alpha'}$ must also be quadratic,
\begin{equation}
V_{\alpha'}(X) = \frac{1}{2}X^\dagger G^{(\alpha')} X - \frac{1}{2}\mathrm{Tr}(G^{(\alpha')}\Sigma).
\label{eq:V}
\end{equation}
By inserting that quadratic form of the quasipotential \eqref{eq:V} in the Poisson equation \eqref{eq:poisson}, we find that
$G$ solves 
\begin{equation}
A^\dagger G^{(\alpha')} + G^{(\alpha')} A = -2F^{(\alpha')}.
\label{eq:G}
\end{equation}
Therefore, the quasistatic thermal response can be written as 
\begin{equation}
R_{\lambda\alpha} = \langle \partial_\lambda V_{\alpha} \rangle_\lambda^s
= -\frac{1}{2}\mathrm{Tr}(G^{(\alpha)} \Phi_{\lambda}),
\label{eq:compactR}
\end{equation}
where $\Phi_\lambda=\partial_\lambda\Sigma$  solves (see \eqref{eq:Sigma})
\begin{equation}
A \Phi_\lambda + \Phi_\lambda A^\dagger  = -[\partial_\lambda D +(\partial _\lambda A)\Sigma +\Sigma (\partial _\lambda A^\dagger)].
\label{eq:Phi}
\end{equation}
Thus, using \eqref{hca}, we get the heat capacity
\begin{equation}
C_{\alpha\alpha'}
=
\langle \partial_{T_\alpha} V_{\alpha'} \rangle
=
-\frac{1}{2}\mathrm{Tr}\!\left(G^{(\alpha')} \Phi_\alpha\right),
\qquad
\alpha,\alpha' \in \{\ell, r\},
\label{eq:Cmatrix}
\end{equation}

\subsection{General results}
One of the first questions is how the heat capacity matrix scales with the system size $N$, which is not evident as the heat-conducting chain exhibits long-range correlations, \cite{Spohn1983}. Interestingly, we find that the leading contribution scales linearly with the system size $C_{\alpha\alpha'} = O(N)$ leading to a well-defined specific heat matrix 
\begin{equation}
c_{\alpha\alpha'} = \frac{1}{N} C_{\alpha\alpha'}
\label{eq:cmatrix}
\end{equation}
also in the thermodynamic limit $N\uparrow\infty$.

\begin{figure}[H]
    \centering
    \begin{subfigure}[t]{0.45\linewidth}
        \centering
        \includegraphics[scale=0.55]{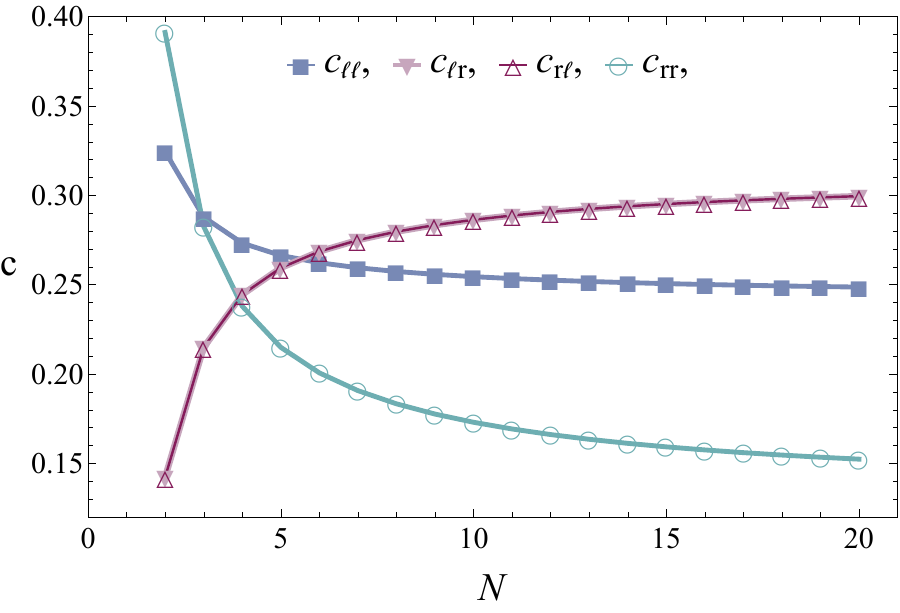}   
        \caption{}
    \end{subfigure}
    \hspace{1cm}
    \begin{subfigure}[t]{0.3\linewidth}
        \centering
        \includegraphics[scale=0.6]{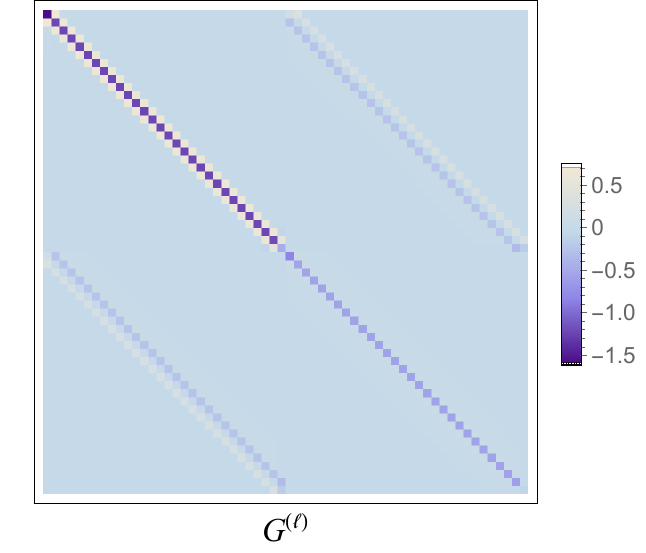}
        \caption{}
    \end{subfigure}
    \caption{(a)Specific heat matrix elements scaling with system size and  (b) the $G$-matrix for $N=30$. The other parameters  for both are $\gamma_\ell=1,\gamma_r=2$, $\kappa=k=1$.}
    \label{fig:extensive}
\end{figure}

That is illustrated in Fig.~\ref{fig:extensive}(a) where $C_{\alpha\alpha'}$ is plotted as a function of $N$.  The reason for the extensivity is the quasilocal structure of the matrices $G^{(\alpha')}$, as illustrated for a chosen parameter set in Fig.~\ref{fig:extensive}(b). \\

A second observation is that for fixed $\gamma_\ell,\gamma_r$, the matrix elements are always independent of the temperatures $T_\ell$ and $T_r$. The reason is that the matrices $A$ and $F$ are temperature-independent, while the noise matrix $D$ depends linearly on the bath temperatures; see \eqref{eq:G} and \eqref{eq:Phi}. Therefore, the nonequilibrium specific heat matrix depends on the driving temperatures only through the dissipation coefficients $\gamma_\ell,\gamma_r$, and hence we expect that for a heat conducting chain the specific heat matrix will in fact depend on temperature(s) through the temperature-dependence of kinetic parameters such as the friction or of the renormalized Hamiltonian (related to the Lamb shift).\\

\begin{figure}[H]
    \centering
    \begin{subfigure}[t]{0.48\textwidth}
        \centering
        \includegraphics[scale=0.55]{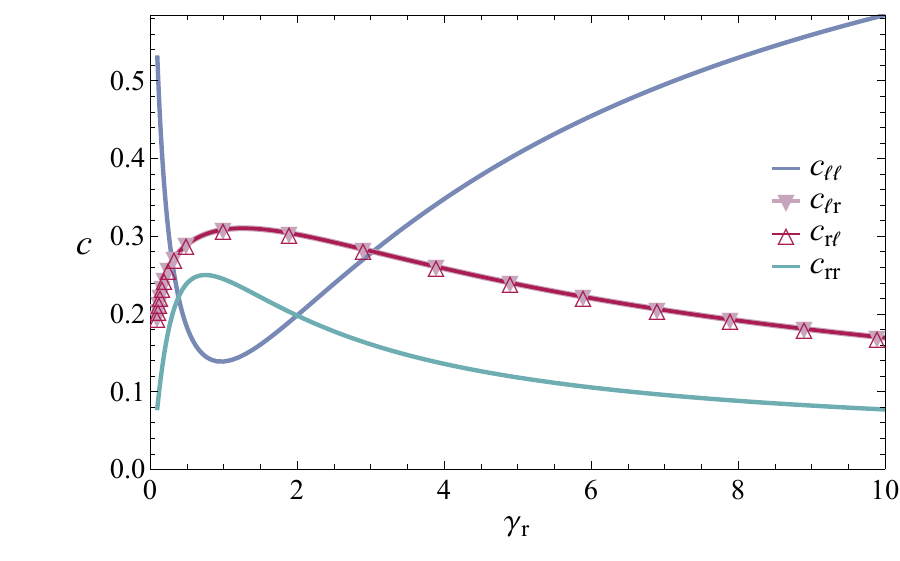}
        \caption{}
    \end{subfigure}
    \hfill
    \begin{subfigure}[t]{0.48\textwidth}
        \centering
        \includegraphics[scale=0.55]{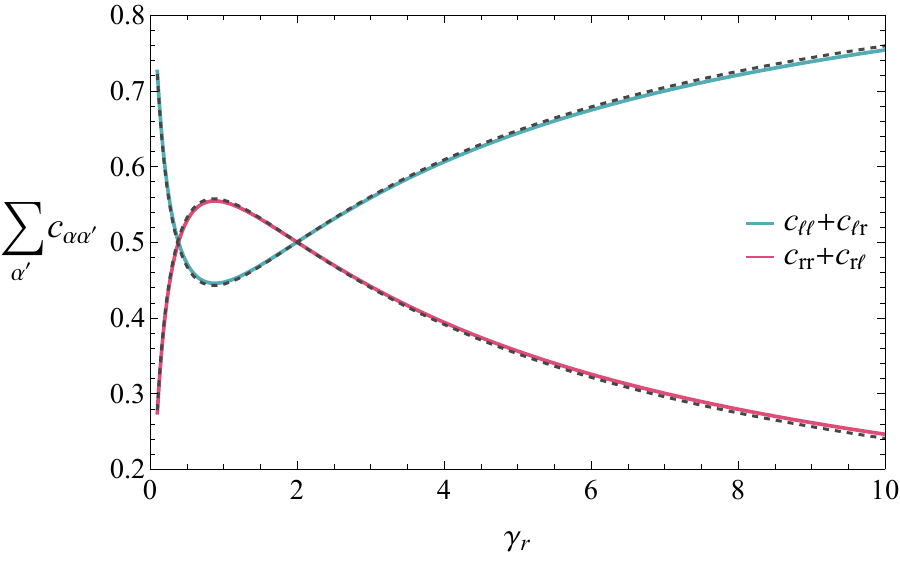}
        \caption{}
    \end{subfigure}
        \caption{(a) Specific heat matrix elements vs $\gamma_r$ and (b) the sum of specific heat elements due to change of the left bath temperature $T_\ell$. The dashed black lines denote the prediction from the bulk temperature profile. The parameters for both figures are $N=60, k=\kappa=1,\gamma_\ell=2$. 
    The colored solid lines denote the row-sums obtained by numerically evaluating the trace formula Eq.~\eqref{eq:Cmatrix}, while the dashed solid lines denote Eq.~\eqref{eq:rowsum}. 
    }
        \label{fig:normalbaths}
    \end{figure}

Fig.~\ref{fig:normalbaths}(a) shows the dependence of the individual matrix elements $c_{\alpha\alpha'}=C_{\alpha\alpha'}/N$ on the boundary friction $\gamma_r$, keeping $\gamma_\ell$ fixed. The diagonal entries show a non-monotonic behavior with $\gamma_r$. For example, $c_{\ell\ell}$, associated with the excess heat from  the left reservoir with a variation of $T_\ell$, initially decreases with an increase in $\gamma_r$ reaches a minimum value, 
beyond which it increases again. This is because for small $\gamma_r$, the right bath is weakly coupled and most of the injected energy is dissipated at the left. Increasing $\gamma_r$ allows energy to flow to the right reservoir, reducing the dissipation at the left. However, for larger values of $\gamma_r$,
the right boundary becomes strongly damped, which hinders transport and leads again to enhanced dissipation at the left. Evidently, this is also reflected in the non-monotonic behavior of $c_{\ell r}$, which shows an opposite trend. The trend in the other two components $c_{rr}$ and $c_{r\ell}$ can also be understood in a same way.

Finally, it is interesting to look at the sum of the specific heat capacity elements due to a change in one of the (say left) bath temperatures, i.e., $c_{\ell}=c_{\ell\ell}+c_{\ell r}$. It measures the total specific heat associated with varying $T_\ell$, irrespective of the reservoir through which the excess heat is dissipated. It turns out that this admits a simple interpretation in terms of the stationary bulk kinetic temperature given in Eq.~\eqref{eq:bulkT}. We find $c_\ell=\partial_{T_\ell} T_\text{bulk}$, in the thermodynamic limit, 
\begin{equation}
c_{\ell}=\frac{\gamma_r}{\gamma_r+\gamma_\ell}+\frac{(k/\kappa)^{2}(\gamma_\ell-\gamma_r)}{(\gamma_r+\gamma_\ell)(1+4k\gamma_r\gamma_\ell/\kappa^2)^{1/2}},\qquad c_r=c_{\ell}\big|_{\gamma_\ell\leftrightarrow\gamma_r}
\label{eq:rowsum}
\end{equation}
See also Fig.~\ref{fig:normalbaths} (b). That reduces to the equilibrium value $1/2$ for $\gamma_\ell=\gamma_r$.

\subsection{Soft-matter baths}\label{sec:sm}
When the Langevin baths arise from coarse-graining more complex environments, such as particle reservoirs in the context of soft matter interactions, the effective frictions $\gamma_{\ell,r}$ may acquire a temperature dependence. It was recently shown in ~\cite{krekels2026negative} that such coarse-grained particle reservoirs can produce a friction coefficient that decreases with temperature as $T^{-2}$, leading to a suppression of transport at higher temperatures as a purely kinetic effect. To capture this thermokinetic effect in a minimal way, we consider a temperature-dependent friction of the form $\gamma_\ell(T_\ell) = \gamma_0 \left(\frac{T_0}{T_\ell}\right)^2$ where $\gamma_0$ sets the coupling strength and $T_0$ is a reference temperature.

\begin{figure}[H]
    \centering
    \begin{subfigure}[t]{0.48\textwidth}
        \includegraphics[scale=0.55]{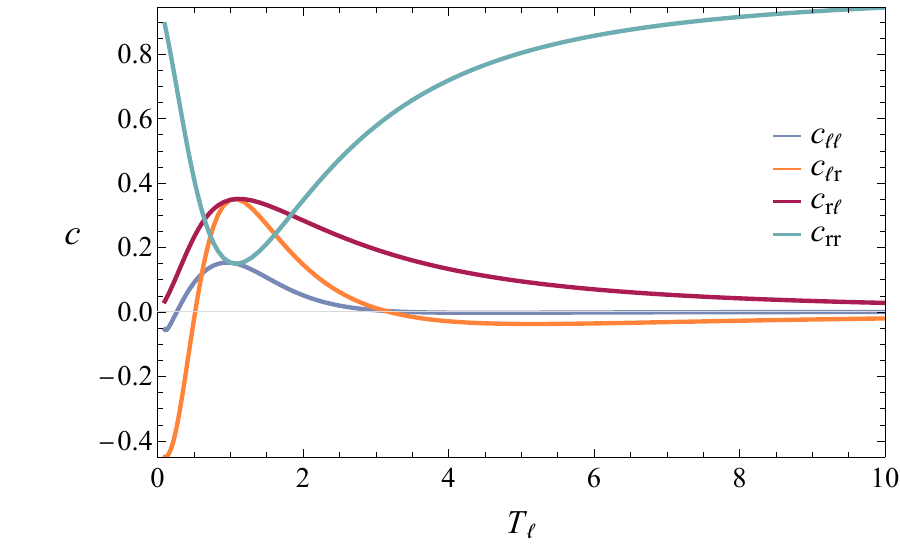}
        \caption{}
    \end{subfigure}
    \hfill
    \begin{subfigure}[t]{0.48\textwidth}
        \includegraphics[scale=0.55]{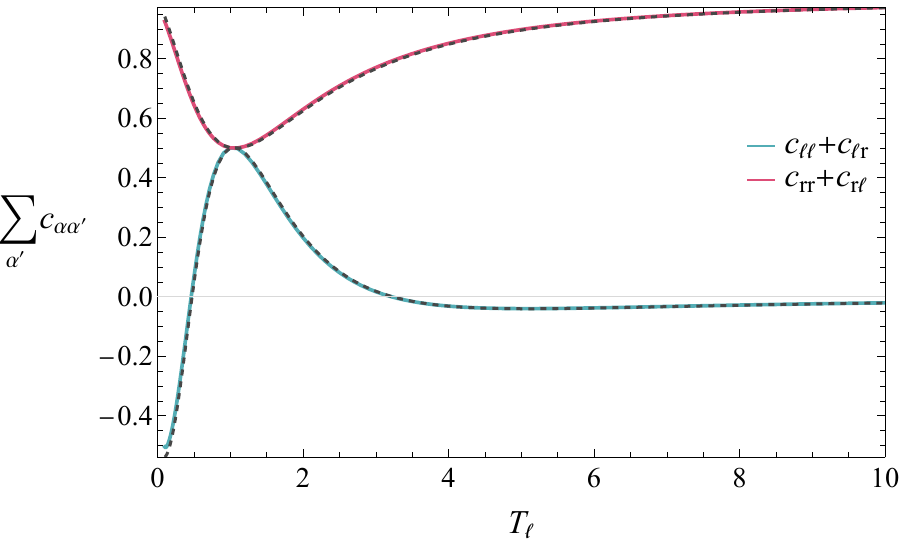}
        \caption{}
    \end{subfigure}
    \caption{(a) Specific heat matrix elements {\it vs} $T_\ell$  and  (b) the sum of specific heat elements due to change of the left bath. The dashed black lines denote the prediction from the bulk temperature profile. The parameters for both figures are  $N=30,\gamma_r=1,T_r=1 $ and $k=\kappa=1$. The colored solid lines denote the row-sums obtained by numerically evaluating the trace formula, while the dashed solid lines denote the specific heat obtained from the bulk energy Eq.~\eqref{eq:bulkT}.}
    \label{fig:softmatter}
    \end{figure}

In Fig.~\ref{fig:softmatter}(a) we show the specific heat matrix elements for this soft-matter bath. In contrast to the usual case, all components now exhibit a pronounced dependence on $T_\ell$, reflecting the explicit temperature dependence of the drift matrix. Interestingly, the specific heat elements associated with variations of the left bath temperature show nontrivial features. The off-diagonal element $c_{\ell r}$ is negative near $T_\ell\to 0$, where the dissipation at the left bath diverges, and becomes negative again for large $T_\ell$, where the coupling effectively vanishes. The corresponding diagonal element $c_{\ell\ell}$ is positive near $T_\ell\to 0$, but decreases and approaches zero at large $T_\ell$, reflecting again the zero dissipation at the left bath. This behavior also carries over to the net specific heat associated with variations of the left reservoir as shown in Fig.~\ref{fig:softmatter}(b). This indicates that increasing the temperature of one reservoir can reduce the excess heat dissipated into a given bath. Such behavior has no analogue in equilibrium calorimetry and highlights the thermokinetic effect. In contrast, the components $c_{r\ell}$ and $c_{rr}$ show a much weaker dependence, as the right bath remains linearly coupled and the quasipotential retains its simple temperature dependence with respect to $T_r$.

\subsection{Quasistatic heat response due to mechanical perturbation}\label{sec:mech}
We can also consider the quasistatic heat response when we make a mechanical perturbation in the chain itself. We realize this by taking  the spring constant $k$ of the chain as the slowly changing parameter $\lambda$. The corresponding response is  $R_{k\alpha'} = \left\langle \partial_k V_{\alpha'} \right\rangle$ with $\alpha'=\ell,r$. 

\begin{figure}[H]
    \centering
    \begin{subfigure}[t]{0.48\textwidth}
        \includegraphics[scale=0.55]{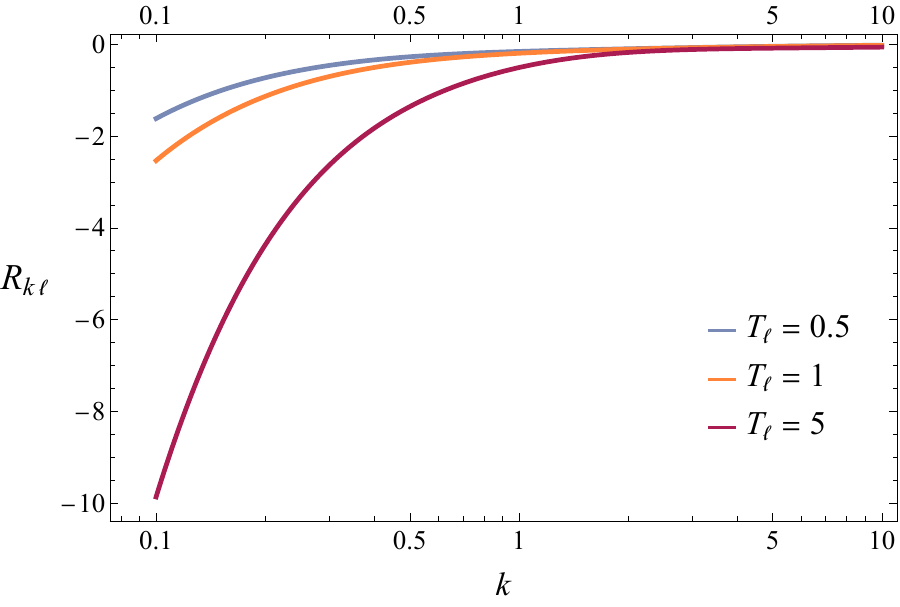}
        \caption{}
    \end{subfigure}
    \hfill
    \begin{subfigure}[t]{0.48\textwidth}
        \includegraphics[scale=0.55]{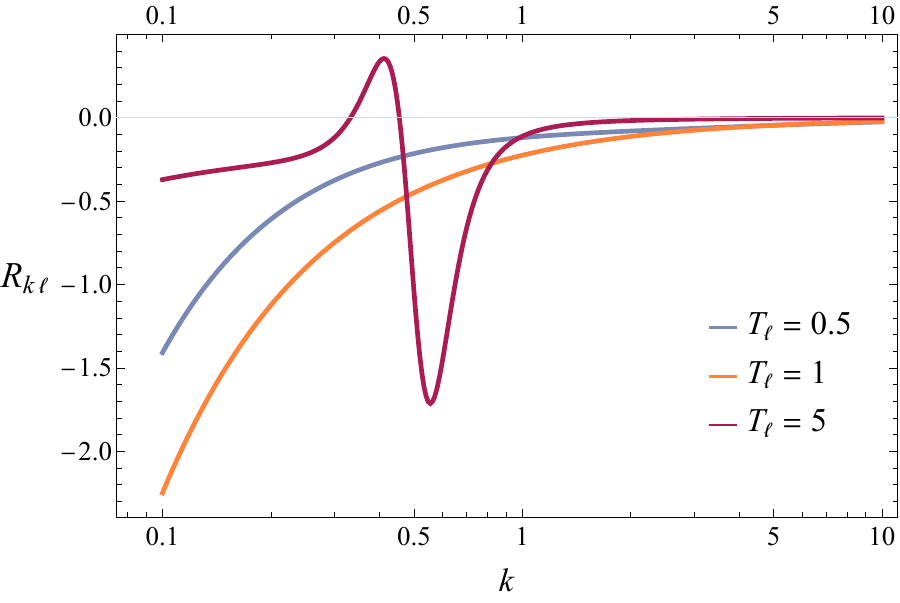}
        \caption{}
    \end{subfigure}
    \caption{Quasistatic heat response to mechanical perturbation $k$ (a) for the baths with constant friction $\gamma_\ell=2,\gamma_r=1$ and  (b) for the soft matter baths $\gamma_\ell\sim\frac{1}{T_\ell^2}$ and $\gamma_r=1$. The parameters  used are $N=20,  k=\kappa=1 $ and $ T_r=1$. }
    \label{fig:mechanical}
    \end{figure}

Fig.~\ref{fig:mechanical}(a) shows the response $R_{k\ell}$ as a function of $k$ for different $T_\ell$. We find that the magnitude of the response is large at small $k$ and decreases with increasing $k$, eventually going to zero. At the same time, the response is predominantly negative, indicating that increasing the stiffness reduces the excess heat dissipated into the reservoirs. Fig.~\ref{fig:mechanical}(b) plots the same for soft-matter baths, and shows an even richer behavior. These nonequilibrium effects are in sharp contrast to the specific heat of an equilibrium harmonic chain, which is independent of the spring constant.

\section{Conclusion}\label{sec:conc}
We extend the notion of nonequilibrium specific heat to a boundary-driven spatially extended system by studying a harmonic chain coupled to two thermal reservoirs at different temperatures, a paradigmatic setup sustaining a steady heat current. In particular, the heat capacity scales with the size of the system.  Using a quasistatic response framework, we compute exactly the nonequilibrium thermodynamic specific heat matrix, which quantifies the excess heat currents generated by slow variations of the bath temperatures. \\
The specific heat is independent of the bath temperatures but depends  on the difference in boundary dissipation coefficients, revealing a thermokinetic contribution. Naturally, when these coefficients acquire temperature dependence, as in the case of soft-matter/particle reservoirs, the specific heat acquires a non-trivial dependence on temperature, and can even be  negative. We also obtain the quasistatic response to a mechanical change in the chain structure.\\
The results suggest a possible way to extend the Dulong-Petit law to boundary-driven molecular gases.

\vspace{1cm}
\section*{Acknowledgments}
  FK is supported by the Research Foundation--Flanders (FWO) postdoctoral fellowship 1232926N.  IS acknowledges funding from the European Union's
Horizon 2024 research and innovation programme
under the Marie Sklodowska - Curie (HORIZONTMAMSCA-PF-EF) grant agreement No. 101205210.
\newpage
\bibliographystyle{unsrt}  
\bibliography{refs-f}

@article{Debye1912,
  author  = {Debye, P.},
  title   = {Zur Theorie der spezifischen W\"arme},
  journal = {Annalen der Physik},
  volume  = {39},
  pages   = {789--839},
  year    = {1912}
}

@article{DulongPetit1819,
  author  = {Petit, A. T. and Dulong, P. L.},
  title   = {Recherches sur quelques points importants de la th{\'e}orie de la chaleur},
  journal = {Annales de Chimie et de Physique},
  volume  = {10},
  pages   = {395--413},
  year    = {1819}
}

@article{Einstein1907,
  author  = {Einstein, A.},
  title   = {{Die Plancksche Theorie der Strahlung und die Theorie der spezifischen W\"arme}},
  journal = {Annalen der Physik},
  volume  = {34},
  pages   = {1--14},
  year    = {1907}
}

@article{Monasterio2019,
  author  = {Mej{\'i}a-Monasterio, C. and Politi, A. and Rondoni, L.},
  title   = {Heat flux in one-dimensional systems},
  journal = {Physical Review E},
  volume  = {100},
  number  = {3},
  pages   = {032139},
  year    = {2019},
  doi     = {10.1103/PhysRevE.100.032139}
}

@article{Spohn1983,
  author  = {Spohn, H.},
  title   = {Long range correlations for stochastic lattice gases in a nonequilibrium steady state},
  journal = {Journal of Physics A: Mathematical and General},
  volume  = {16},
  number  = {18},
  pages   = {4275--4291},
  year    = {1983}
}

@article{aoki2006energy,
  author  = {Aoki, K. and Lukkarinen, J. and Spohn, H.},
  title   = {Energy transport in weakly anharmonic chains},
  journal = {Journal of Statistical Physics},
  volume  = {124},
  number  = {5},
  pages   = {1105--1129},
  year    = {2006}
}

@article{berryo,
  author  = {Beyen, A. and Khodabandehlou, F. and Maes, C.},
  title   = {Quasistatic response for nonequilibrium processes: evaluating the {B}erry potential and curvature},
  journal = {arXiv},
  volume  = {2512.01654},
  year    = {2025},
  eprint  = {2512.01654},
  archivePrefix = {arXiv},
  primaryClass  = {cond-mat.stat-mech},
  url     = {https://arxiv.org/abs/2512.01654}
}

@article{calo,
  author  = {Maes, C. and Neto{\v{c}}n{\'y}, K.},
  title   = {Nonequilibrium calorimetry},
  journal = {Journal of Statistical Mechanics: Theory and Experiment},
  volume  = {2019},
  number  = {11},
  pages   = {114004},
  year    = {2019},
  doi     = {10.1088/1742-5468/ab4589}
}

@article{cejp,
  author  = {Pe{\v{s}}ek, J. and Boksenbojm, E. and Neto{\v{c}}n{\'y}, K.},
  title   = {Model study on steady heat capacity in driven stochastic systems},
  journal = {Open Physics},
  volume  = {10},
  number  = {3},
  pages   = {692--701},
  year    = {2012},
  doi     = {10.2478/s11534-012-0053-8}
}

@article{dhar2001heat,
  author  = {Dhar, A.},
  title   = {Heat conduction in the disordered harmonic chain revisited},
  journal = {Physical Review Letters},
  volume  = {86},
  number  = {26},
  pages   = {5882},
  year    = {2001}
}

@article{dhar2006heat,
  author  = {Dhar, A. and Roy, D.},
  title   = {Heat transport in harmonic lattices},
  journal = {Journal of Statistical Physics},
  volume  = {125},
  number  = {4},
  pages   = {801--820},
  year    = {2006}
}

@article{dhar2008heat,
  author  = {Dhar, A.},
  title   = {Heat transport in low-dimensional systems},
  journal = {Advances in Physics},
  volume  = {57},
  number  = {5},
  pages   = {457--537},
  year    = {2008}
}

@article{epl,
  author  = {Boksenbojm, E. and Maes, C. and Neto{\v{c}}n{\'y}, K. and Pe{\v{s}}ek, J.},
  title   = {Heat capacity in nonequilibrium steady states},
  journal = {Europhysics Letters},
  volume  = {96},
  number  = {4},
  pages   = {40001},
  year    = {2011},
  doi     = {10.1209/0295-5075/96/40001}
}

@article{iannacchione1992calorimetric,
  author  = {Iannacchione, G. S. and Finotello, D.},
  title   = {Calorimetric study of phase transitions in confined liquid crystals},
  journal = {Physical Review Letters},
  volume  = {69},
  number  = {14},
  pages   = {2094},
  year    = {1992}
}

@article{jchemphys,
  author  = {Khodabandehlou, F. and Maes, C. and Neto{\v{c}}n{\'y}, K.},
  title   = {A {N}ernst heat theorem for nonequilibrium jump processes},
  journal = {The Journal of Chemical Physics},
  volume  = {158},
  number  = {20},
  year    = {2023},
  doi     = {10.1063/5.0142694}
}

@phdthesis{jir,
  author = {Pe{\v{s}}ek, J.},
  title  = {Heat Processes in Non-Equilibrium Stochastic Systems},
  school = {Charles University in Prague},
  year   = {2014}
}

@book{kittel2018introduction,
  author    = {Kittel, C. and McEuen, P.},
  title     = {Introduction to Solid State Physics},
  publisher = {John Wiley \& Sons},
  year      = {2018}
}

@article{krekels2026negative,
  author  = {Krekels, S. and Maes, C. and Santra, I. and Zhai, R.},
  title   = {Negative Differential Heat Conductivity in a Harmonic Chain Coupled to a Particle Reservoir},
  journal = {arXiv},
  year    = {2026},
  eprint  = {2604.00777},
  archivePrefix = {arXiv},
  url     = {https://arxiv.org/abs/2604.00777}
}

@article{lepri2003thermal,
  author  = {Lepri, S. and Livi, R. and Politi, A.},
  title   = {Thermal conduction in classical low-dimensional lattices},
  journal = {Physics Reports},
  volume  = {377},
  number  = {1},
  pages   = {1--80},
  year    = {2003}
}

@article{lepri2005studies,
  author  = {Lepri, S. and Livi, R. and Politi, A.},
  title   = {Studies of thermal conductivity in {F}ermi--{P}asta--{U}lam-like lattices},
  journal = {Chaos: An Interdisciplinary Journal of Nonlinear Science},
  volume  = {15},
  number  = {1},
  year    = {2005}
}

@article{lukkarinen2008anomalous,
  author  = {Lukkarinen, J. and Spohn, H.},
  title   = {Anomalous energy transport in the {FPU}-$\beta$ chain},
  journal = {Communications on Pure and Applied Mathematics},
  volume  = {61},
  number  = {12},
  pages   = {1753--1786},
  year    = {2008}
}

@article{maes2003heat,
  author  = {Maes, C. and Neto{\v{c}}n{\`y}, K. and Verschuere, M.},
  title   = {Heat conduction networks},
  journal = {Journal of Statistical Physics},
  volume  = {111},
  number  = {5},
  pages   = {1219--1244},
  year    = {2003}
}

@article{mathnernst,
  author  = {Khodabandehlou, F. and Maes, C. and Maes, I. and Neto{\v{c}}n{\'y}, K.},
  title   = {The vanishing of excess heat for nonequilibrium processes reaching zero ambient temperature},
  journal = {Annales Henri Poincar{\'e}},
  year    = {2023},
  issn    = {1424-0661},
  doi     = {10.1007/s00023-023-01367-1}
}

@article{nakazawa1970lattice,
  author  = {Nakazawa, H.},
  title   = {On the lattice thermal conduction},
  journal = {Progress of Theoretical Physics Supplement},
  volume  = {45},
  pages   = {231--262},
  year    = {1970}
}

@article{pois,
  author  = {Khodabandehlou, F. and Maes, C. and Neto{\v{c}}n{\'y}, K.},
  title   = {On the {P}oisson equation for nonreversible {M}arkov jump processes},
  journal = {Journal of Mathematical Physics},
  volume  = {65},
  number  = {4},
  year    = {2024},
  doi     = {10.1063/5.0184909}
}

@article{rieder1967properties,
  author  = {Rieder, Z. and Lebowitz, J. L. and Lieb, E.},
  title   = {Properties of a harmonic crystal in a stationary nonequilibrium state},
  journal = {Journal of Mathematical Physics},
  volume  = {8},
  number  = {5},
  pages   = {1073--1078},
  year    = {1967}
}

@article{santra2022activity,
  author  = {Santra, I. and Basu, U.},
  title   = {Activity driven transport in harmonic chains},
  journal = {SciPost Physics},
  volume  = {13},
  number  = {2},
  pages   = {041},
  year    = {2022}
}

@article{wunderlich1997heat,
  author  = {Wunderlich, B.},
  title   = {The heat capacity of polymers},
  journal = {Thermochimica Acta},
  volume  = {300},
  number  = {1--2},
  pages   = {43--65},
  year    = {1997}
}

@book{yeomans1992statistical,
  author    = {Yeomans, J. M.},
  title     = {Statistical Mechanics of Phase Transitions},
  publisher = {Clarendon Press},
  year      = {1992}
}
\onecolumngrid
\end{document}